\documentclass[10pt]{article}
\usepackage{wrapfig}

\usepackage[english]{babel}

\usepackage[letterpaper,top=1in,bottom=1in,left=1in,right=1in,marginparwidth=1.75cm]{geometry}

\usepackage{amsmath}
\usepackage{graphicx}
\usepackage[colorlinks=true, allcolors=blue]{hyperref}
\usepackage{authblk}
\usepackage{parskip}
\usepackage[font=small]{caption}

\title{\textbf{\Large FOXDEN: FAIR Services for AI-Ready Scientific Datasets}}
\author[1]{Valentin Kuznetsov}
\author[1]{Werner M. Sun}
\author[1]{Keara Soloway}
\author[1]{Rolf Verberg}
\author[1]{Katherine S. Shanks}
\author[1]{Suchismita Sarker}
\author[2]{David S. Butcher} %https://orcid.org/0000-0002-1268-4150
\author[1]{Kelly E. Nygren}
\affil[1]{Cornell High Energy Synchrotron Source, Cornell University, Ithaca, NY 14853}
\affil[2]{National High Magnetic Field Laboratory, Tallahassee, FL 32310}

\date{September 2, 2026}

\begin{document}
\maketitle

\begin{abstract}
Scientific datasets are most compatible with AI workflows when they are described by rich, machine-readable metadata and provenance information, following the FAIR guiding principles. The FAIR Open-Science Extensible Data Exchange Network (FOXDEN) is a set of cyberinfrastructure building blocks developed at the Cornell High Energy Synchrotron Source (CHESS) for annotating raw, reduced, and analyzed datasets with both structured and unstructured metadata and provenance records. It also allows researchers to publish these records with Digital Object Identifiers (DOIs) to create AI-ready datasets. We describe FOXDEN's architecture and its deployment at CHESS and at the National High Magnetic Field Laboratory, where it serves as a foundation for future agentic scientific workflows.
\end{abstract}

\section{Introduction}
\label{sec:intro}

The recent emergence of artificial intelligence (AI) in scientific research has highlighted the urgent need for high-quality experimental datasets for training and validating AI models. In order for such datasets to be discovered and responsibly reused, either by AI agents or human scientists, they must first be described with rich metadata that identifies their content and provides context for their use~\cite{metadata1,metadata2,metadata3,metadata4,haberman-reuse}. Furthermore, establishing trust in AI requires governance and risk management processes built around the provenance of training datasets~\cite{provenance1,sperhac_2026_19892854,provenance2,longpre2023dataprovenanceinitiativelarge}.
Therefore, the dissemination of machine-readable metadata and provenance---in compliance with the FAIR guiding principles (findable, accessible, interoperable, reusable)~\cite{fair-wilkinson}---plays a role of newfound importance in data-intensive science.

At the Cornell High Energy Synchrotron Source (CHESS), we have developed the FAIR Open-Science Extensible Data Exchange Network (FOXDEN), a suite of lightweight, modular data services that allows researchers to easily record metadata, provenance, and other research artifacts in real time to accompany their experimental datasets. CHESS's current Data Policy essentially archives all raw and processed data forever. FOXDEN is specifically designed to handle these large, unportable datasets and heterogeneous use cases. It helps scientists turn their research artifacts into annotated, AI-ready datasets and publish them with Digital Object Identifiers (DOIs) according to FAIR principles.

Although FOXDEN originated at CHESS, it is designed to be deployed by any user facility or research group. CHESS is an ideal environment for developing such interdisciplinary cyberinfrastructure because it serves a wide range of users from different scientific domains, across multiple beamlines that each specialize in different experimental techniques. Experimental modalities at CHESS range from diffraction techniques (like high-energy diffraction microscopy of structural materials, high-throughput high-dynamic range mapping of quantum materials, or macromolecular crystallography for structural biology) to imaging and spectroscopy with x-ray tomography, fluorescence, and absorption. CHESS beamlines employ numerous data collection workflows and metadata practices, and there is wide variation in data volumes, with some beamlines collecting 200 TB of raw and processed data per year, while others collect less than 5 TB. Research workflows extend across multiple platforms, starting with data collection through centralized CHESS resources, followed by data analysis on a personal laptop or CHESS HPC clusters, using either beamline-written or third-party software. Therefore, CHESS researchers produce metadata through many mechanisms, and the metadata itself can be structured (\emph{e.g.}, detector parameters) or unstructured (\emph{e.g.}, freeform notes), with both high- and low-volume streams.

The scientific requirement to support multi-platform FAIR workflows in heterogeneous environments with diverse forms of metadata has led to the following architectural features in FOXDEN:
\begin{itemize}
    \item Modular design to allow FOXDEN services to match the shape of any experimental workflow.
    \item Use of common, non-proprietary protocols to enable seamless integration with existing and emerging data collection and data analysis frameworks.
    \item Native capabilities for curating, publishing, and disseminating metadata.
    \item Customizable per-beamline or per-user metadata schemas with unified views provided where possible.
    \item Support for community metadata standards without strict enforcement.
    \item Segmentation of backend repositories for efficient management of data and metadata collections with unequal volumes.
    \item Generalized, abstracted interfaces for the underlying storage and database resources to facilitate streamlined deployment in other environments.
    \item Support for granular, per-record permissions to view, create, and edit metadata records based on a user's group membership.
\end{itemize}
With these features, FOXDEN provides a unified but extensible framework for reproducible science and cross‑facility collaboration without imposing a monolithic data model on the scientific community.

\section{Research Workflow}
\label{sec:research-workflow}

\begin{figure}[ht]
\begin{center}
\includegraphics[width=\linewidth]{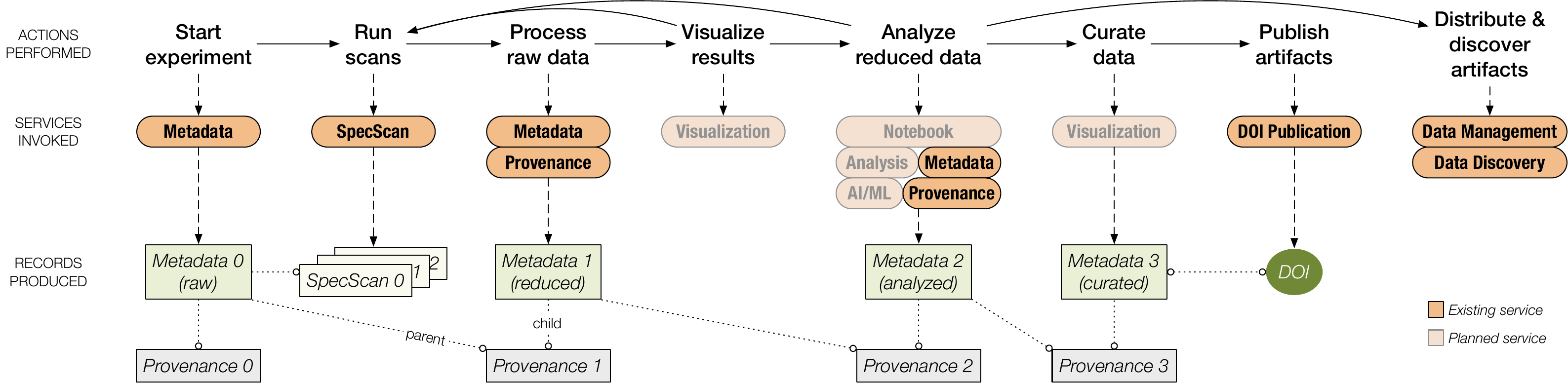}
\caption{\label{fig:foxden-workflow} A typical CHESS research workflow. FOXDEN services are invoked at each step to produce chains of Metadata, Provenance, and SpecScan records that can be published under a single DOI. Dotted lines connect linked records, and an open circle on one end denotes a reference to the record at the other end.}
\end{center}
\end{figure}

FOXDEN services are web applications written in Go, built on backend database repositories, with APIs for command-line interaction. Each service performs a different function in the data ecosystem: metadata management, provenance tracking, publication, etc. Figure~\ref{fig:foxden-workflow} shows a typical CHESS research workflow. At every step, FOXDEN services can be invoked (by either a researcher or an automated software process) to create three types of records: \textbf{Metadata} for datasets, \textbf{Provenance} for parent/child relationships between datasets, and \textbf{SpecScan} records for sub-datasets.

Metadata and Provenance records are completely generic, while SpecScan records are tailored to the SPEC~\cite{spec} data acquisition (DAQ) program used at CHESS. However, FOXDEN does not depend on SPEC for any of its functions. Indeed, FOXDEN is designed for use with any DAQ system, or even none at all. For DAQ systems that produce topologies not shown in Figure~\ref{fig:foxden-workflow}, we can introduce additional record types and services that encode those inter-record relationships.

As a CHESS experiment progresses, chains of FOXDEN records are assembled in parallel with the data products they describe. Metadata records may contain information about the facility and instruments used to collect the dataset, descriptions of the samples being studied, experimental parameters, and data analysis results (if any). To form the chains, FOXDEN links Metadata records together using Provenance records.

A raw dataset at CHESS often consists of multiple scans, in which a sample is studied under varying conditions (\emph{e.g.}, position or temperature). These scans are orchestrated by SPEC, and SpecScan records contain metadata related to individual scans. As illustrated in Figure~\ref{fig:foxden-workflow}, SpecScan records are directly associated with their parent Metadata record without an intermediate Provenance record.

After they are created, Metadata records can be amended and/or updated with additional notes and attributes, which aids with their integration into external repositories or data services. Critically, researchers might perform different portions of this workflow using a variety of independent systems, but when metadata from these systems are deposited in FOXDEN, they are aggregated into a single, unified representation of the sequence of events.
At any point in the workflow, researchers may choose to omit Metadata records and their associated Provenance records without affecting the integrity of the chains. At the end of the workflow, researchers can publish their records by minting a Digital Object Identifier (DOI), either for the entire collection or only for selected segments.

By giving researchers a means to simultaneously record data and metadata, FOXDEN lowers the barriers to complying with FAIR principles---metadata stored in FOXDEN can be published almost immediately with only minimal curation. Thus, FOXDEN promotes efficiency, transparency, and reproducibility in science.

\section{Metadata Engineering}
\label{sec:metadata-engineering}

Based on the requirements outlined in Section~\ref{sec:intro}, we have implemented the following high-level features in FOXDEN's metadata handling services:

\begin{enumerate}
    \item \emph{Metadata is decoupled from the data that it describes.} In data-intensive experiments at CHESS, raw datasets can be as large as 100 TB, while their associated metadata is less than 1 MB. So, in general, metadata is easily portable, while the data itself is not. Because of their size differences, data and metadata require separate infrastructure and strategies for storage, management, and dissemination. At CHESS, data files have long been kept on dedicated storage devices and archival tape libraries. FOXDEN augments this existing solution with lightweight metadata capabilities.
    \item \emph{Multiple internal schemas are supported.} By allowing researchers to define schemas on a per-beamline and per-application basis, we support a variety of use cases, as described in Section~\ref{sec:intro}. For further flexibility, 
    we intentionally decouple FOXDEN's internal schemas from community metadata standards, thus insulating them from specific conventions that may evolve over time. FOXDEN metadata may be converted into standards-compliant formats using a translation service (see Section~\ref{sec:info-layout}). We also decouple the metadata's representation from its storage format (JSON, SQL tables, etc.) so we can render the records dynamically using the best format for each use case.
    \item \emph{Metadata records are modular and independent.} In the decentralized architecture described in Section~\ref{sec:architecture}, services operate independently, so the types of records that they handle must also be structurally independent. For example, SQL tables contain no foreign keys. Instead, every record is labeled with a globally unique dataset identifier (DID) that cross-references records within a chain (see Section~\ref{sec:did}). Because these links are dynamically generated, not prescribed by a schema, FOXDEN gives researchers the flexibility to document workflows of any arbitrary shape.
\end{enumerate}

\subsection{Types of Metadata Records}
\label{sec:record-types}

FOXDEN currently handles three types of experimental metadata that were introduced in Section~\ref{sec:research-workflow} and are described more fully below.

\subsubsection{Generic Metadata}
\label{sec:generic-metadata}
Generic Metadata records (denoted with a capital M) are intended primarily for datasets, but they can also be used to represent other entities, such as facilities and instruments. Per-beamline schemas are defined by beamline scientists in JSON format, as shown in Figure~\ref{fig:example-schema}. Allowed types for attributes include basic types like \texttt{bool}, \texttt{int}, \texttt{float}, \texttt{string}; lists and \texttt{struct}s composed of these basic types; as well as lists of \texttt{struct}s. Attributes can be designated as required or optional, but every schema must include a DID attribute. Descriptions and physical units (\emph{e.g.}, keV, mm) may also be included as parts of attributes.

To prevent inconsistencies across schemas, we have adopted a naming convention in which keys of attributes are written in lower case with underscores to separate words (snake$\_$case). This convention helps search queries to more effectively find and retrieve records across schemas with common attributes.

\begin{wrapfigure}{r}{0.3\textwidth}
\vspace{-0.2cm}
    \begin{center}
{\tiny
\begin{verbatim}
  {
    "key": "buffer",
    "type": "string",
    "optional": true,
    "multiple": false,
    "section": "Sample",
    "description": "Composition and pH of buffer",
    "units": "",
    "placeholder": "50 mM Hepes, pH 7.5"
  },
  {
    "key": "temperature",
    "type": "float64",
    "optional": false,
    "multiple": false,
    "section": "Sample",
    "description": "Temp at sample cell [Celsius]",
    "units": "Celsius",
    "placeholder": "25"
  },
  {
    "key": "column",
    "type": "string",
    "optional": true,
    "multiple": false,
    "section": "Chromatography",
    "description": "Column type",
    "placeholder": "Superdex 200 Increase 10/300"
  }, ...
\end{verbatim}}
\caption{Fragment of an example Metadata schema.}
	\label{fig:example-schema}
    \end{center}
\vspace{-1cm}
\end{wrapfigure}

We provide flexibility for different use cases with the following extensions to Metadata schemas:
\begin{itemize}
	\item \emph{Composable schemas:} As schemas grow in complexity, they can be factorized into reusable sub-schemas. This mechanism allows common base schemas to be combined with domain‑specific blocks, which facilitates the maintenance of multiple heterogeneous schemas while preserving consistency among them.
	\item \emph{Embeddable schemas:} A schema may include a list of \texttt{struct}s as an attribute, where a \texttt{struct} is defined by its own sub-schema. Thus, a record may contain multiple instances of a sub-record that describe, for example, the known crystollagraphic phases in a sample, which can vary in number.
    \item \emph{Ad hoc schema extensions:} Users may update existing Metadata records with freeform notes and file attachments (\emph{e.g.}, plots or figures), as well as additional custom key-value pairs that capture specialized, project‑specific metadata. Users may also submit \emph{ad hoc} records composed entirely of custom key-value pairs.
\end{itemize}
In the future, we will introduce a mechanism to handle schema evolution as well as time-varying records (such as calibrations), where the versions of a record are ordered sequentially by their intervals of validity.

\subsubsection{Provenance}
Provenance is a specialized type of metadata that encodes the links between datasets by identifying the parent and child DIDs. Provenance also records the operating system, software, and Python environment used to produce the child dataset. For raw data, the Provenance includes the DAQ software (SPEC) version and macros used during data collection.

Provenance is treated as a first‑class entity distinct from the Metadata it describes. Provenance records have a fixed schema, but multiple parents are allowed, forming a directed‑acyclic graph rather than a simple linear chain. This flexible topology captures complex relationships, such as composite datasets, as shown in Figure~\ref{fig:info-layout}. This full lineage recorded by FOXDEN is an essential component of trustworthy AI-ready datasets.

\subsubsection{SpecScan}
SpecScan records are another specialized type of metadata with a fixed schema. They complement the dataset-level Metadata records by describing the constituent scan(s) within a dataset. SpecScan records are directly linked to their appropriate Metadata record simply by sharing a value for the DID field. Unlike Metadata records, SpecScan records do not need to have a unique value for DID because a dataset may consist of multiple scans. 

In addition to the DID field, the schema for SpecScan records contains the following attributes: the version of SPEC used to run the scan, the start time of the scan, the command used to run the scan, the location to which SPEC wrote data for the scan, and the positions of all configured SPEC motors when the scan was run. 

SpecScan records also contain a field in which users may include supplementary metadata about that scan as schemaless key-value pairs. For instance, in an experiment where a sample is scanned multiple times under different conditions (\emph{e.g.}, varying temperature, beam energy, applied magnetic field, etc.), users may include the value of that varying parameter in each scan's SpecScan record. They may also add a parameter to indicate the quality of the scan's data to help identify the suitability of that scan's data for inclusion in further analysis.

\subsection{Information Layout and Use in AI Applications}
\label{sec:info-layout}

\begin{figure}[ht]
\begin{center}
\includegraphics[width=0.9\linewidth]{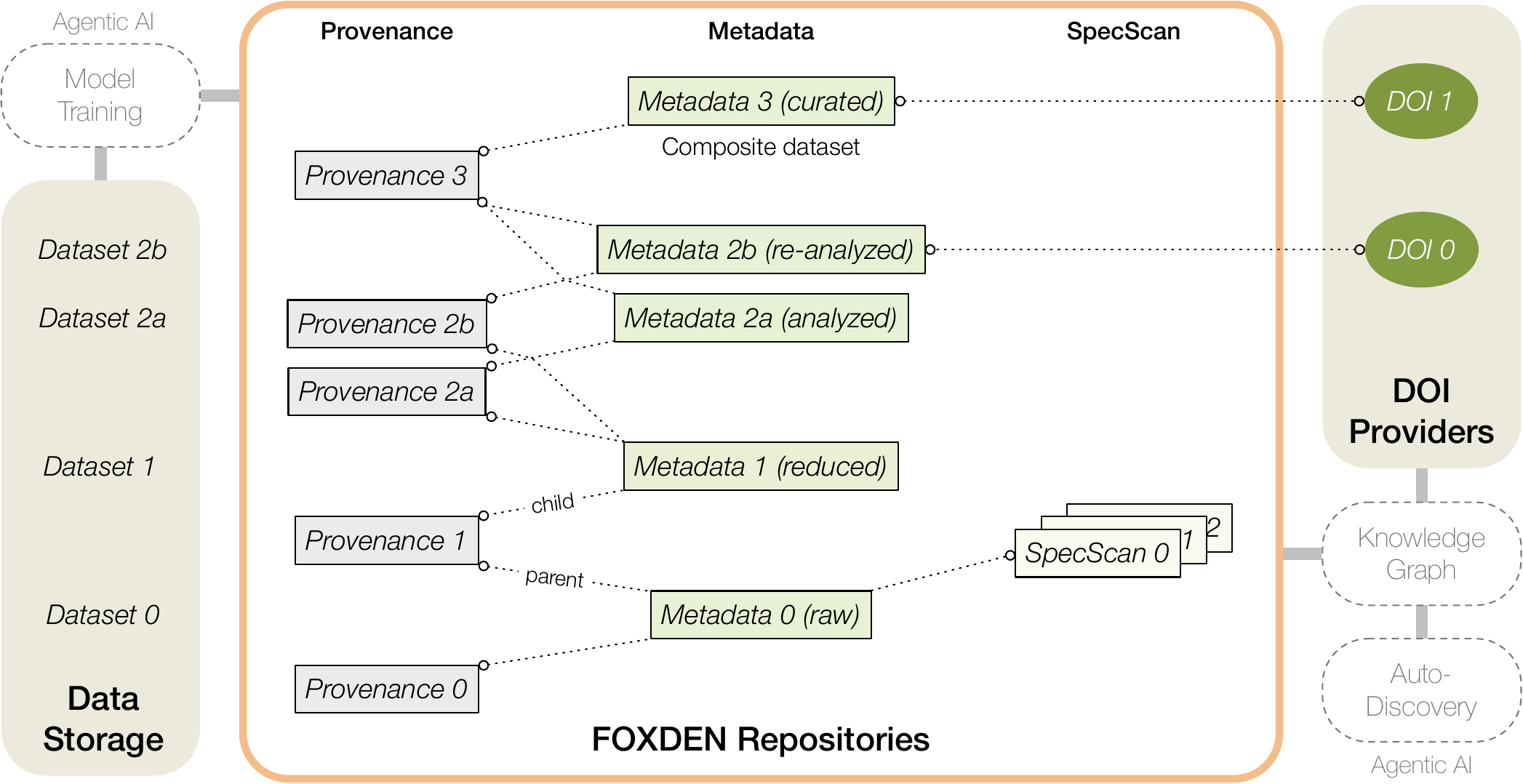}
\caption{\label{fig:info-layout} Relationships among metadata records, stored datasets, and DOIs in the FOXDEN information ecosystem. FOXDEN records are connected by dotted lines, and an open circle on one end denotes a reference to the record at the other end. Interactions with agentic AI components (dashed grey ovals) are also shown schematically.}
\end{center}
\end{figure}

Figure~\ref{fig:info-layout} shows the relationships among entities in the FOXDEN information ecosystem generated by a hypothetical research workflow illustrating numerous topologies. Raw, reduced, and analyzed datasets residing in data storage are associated with FOXDEN Metadata records, accompanied by SpecScan records for the constituent scans in the raw dataset (Dataset 0). Provenance records document the history of each dataset.

In this workflow, after the analyzed dataset (2a) is produced, the researcher decides to redo the analysis using an alternate procedure that produces a second analyzed dataset (2b). The Metadata records for Datasets~2a and 2b are linked to their common parent (through Provenance), resulting in two child datasets for Dataset~1. These records are then published with external DOI providers, as discussed below in Section~\ref{sec:publication-strategy}.

Figure~\ref{fig:info-layout} also shows how FOXDEN plays a role in future AI applications.
Auto-discovery agents can locate datasets through a knowledge graph that translates FOXDEN records into standards-compliant metadata. This graph internally maps and organizes metadata records from multiple sources (not limited to FOXDEN), aggregates and distills this information in response to queries or prompts, and presents the results to agentic workflows in a standard format. Once a dataset is discovered, AI agents can then retrieve it for model training or validation. FOXDEN records are also critical to this step, providing the necessary context for interpreting that dataset, such as: definitions and reference points for measurements, instrument settings, and data quality assessments.
Thus, FOXDEN represents a piece of the foundational data infrastructure that any scalable agentic system will require.

\subsection{Dataset Identifiers}
\label{sec:did}
Dataset identifiers (DIDs) are human-readable, unique identifiers for Metadata records, and they are mandatory attributes of Metadata schemas. By default, DIDs are composed by concatenating a configurable list of schema attributes, but this convention can be overridden when a Metadata record is created.

DIDs are also extensible, which allows them to encode the data processing tier for each dataset. Table~\ref{tab:did} shows example DIDs for the Metadata records shown in Figure~\ref{fig:info-layout}. The raw dataset has a base DID that is subsequently extended for the derived datasets. This base DID connects all of the datasets in the chain, while adding extensions satisfies the uniqueness requirement for each DID.

\begin{table}[ht]
\centering
\footnotesize
\caption{DIDs for Metadata records in Figure~\ref{fig:info-layout}. The raw dataset's DID is extended for the derived datasets.}
\begin{tabular}{cl}
\hline\hline
Dataset & \multicolumn{1}{c}{Example Dataset ID (DID)} \\
\hline
Raw & \texttt{/beamline=3a/btr=test-123-a/cycle=2025-3/sample\_name=test} \\
Reduced & \texttt{/beamline=3a/btr=test-123-a/cycle=2025-3/sample\_name=test/tier=reduced} \\
Analyzed & \texttt{/beamline=3a/btr=test-123-a/cycle=2025-3/sample\_name=test/tier=reduced/user=jane:20260101\_001} \\
Re-analyzed & \texttt{/beamline=3a/btr=test-123-a/cycle=2025-3/sample\_name=test/tier=reduced/user=jane:20260203\_003} \\
Curated & \texttt{/beamline=3a/btr=test-123-a/cycle=2025-3/sample\_name=test/tier=reduced/jane\_composite} \\
\hline\hline
\end{tabular}
\label{tab:did}
\end{table}

\section{DOI Publication Strategy}
\label{sec:publication-strategy}

We adopt a lightweight publication strategy, in which metadata records are published with external DOI providers, but the corresponding datasets remain in storage at CHESS because of their large size. The published metadata contains references to data locations at CHESS, and the DOI landing pages include instructions for requesting data access as well as information about licenses for data reuse (\emph{e.g}, Creative Commons). The original researchers retain control over who is authorized to access and reuse each of their datasets.

FOXDEN supports publishing chained metadata structures to any depth. Researchers may elect to publish an entire metadata structure, including all child records and their Provenance records, or only a sub-structure (\emph{e.g.}, a single record) while keeping the other records in the structure private. FOXDEN automatically assembles the structures using Provenance record information.

Figure~\ref{fig:info-layout} demonstrates two of the many possibilities for metadata publication. The first DOI (0) is minted for the linear chain of records ending with Metadata/Provenance 2b, \emph{i.e.}, 2b $\rightarrow$ 1 $\rightarrow$ 0.
The second DOI (1) is minted after a data curation step, in which a virtual, composite dataset is created from the two analyzed datasets. Metadata 3 is a standalone record whose Provenance has two parents. When it is published, the entire collection of Metadata/Provenance records under Metadata 3 can be captured under a single DOI: 3 $\rightarrow$ 2a+2b $\rightarrow$ 1 $\rightarrow$ 0. This collection may also be truncated for publication.

\section{FOXDEN Architecture}
\label{sec:architecture}

FOXDEN’s service-oriented architecture is built around loosely coupled, interchangeable services, which make the system both modular and adaptable.  Each service exposes its own set of Representational State Transfer (REST) APIs and can be integrated with any data management platform via the standard HTTP protocol. 
Because it is not monolithic, FOXDEN's design enables effortless scaling across multiple geographic sites: each site can select the subset of components it requires, and sites can be synchronized transparently.  

\begin{figure}[ht]
\begin{center}
\includegraphics[width=0.9\linewidth]{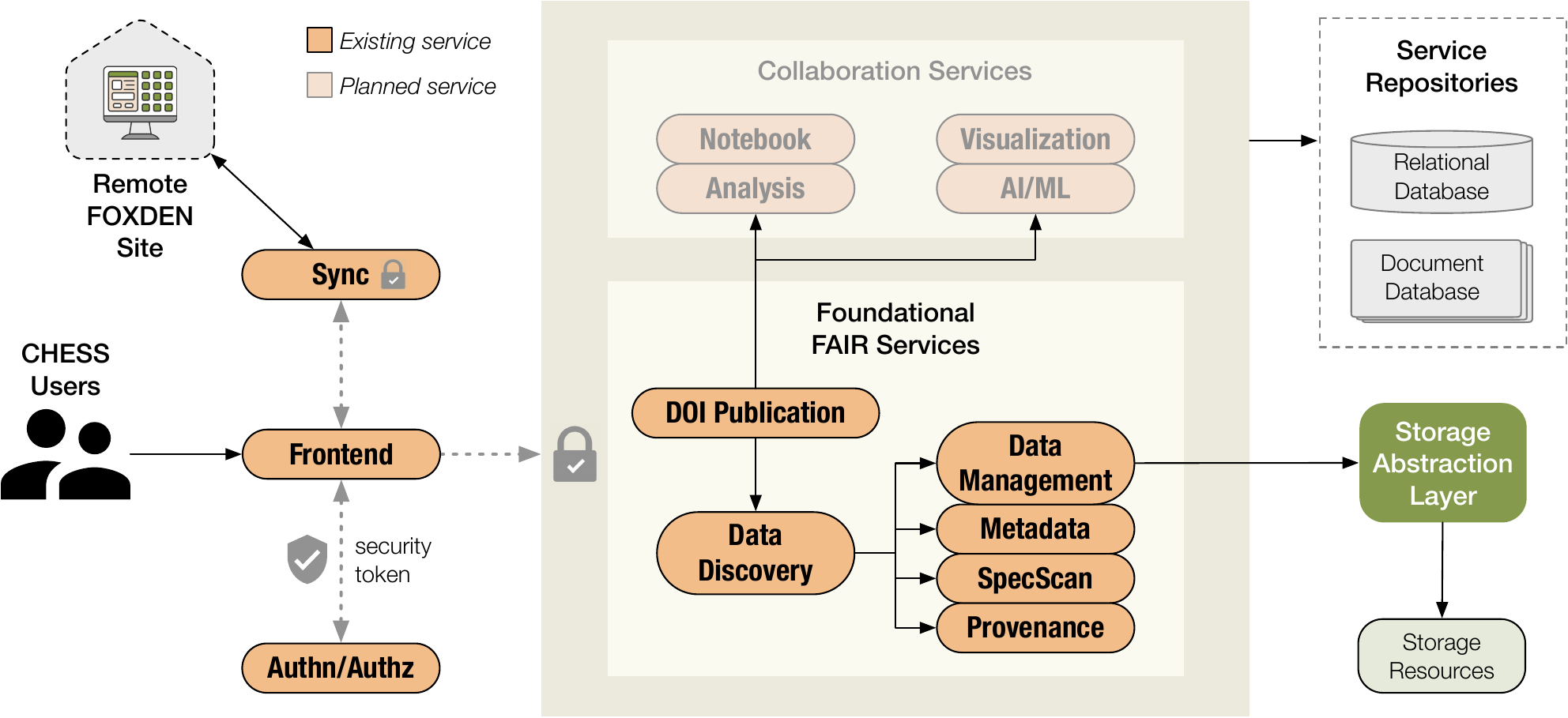}
\caption{\label{fig:foxden_architecture2} Architecture of existing and planned FOXDEN services, repositories, and storage abstraction layer which can be deployed and synchronized across multiple sites. Access to FOXDEN services is restricted with security tokens.}
\end{center}
\end{figure}

The overall layout of the FOXDEN ecosystem is illustrated in Figure~\ref{fig:foxden_architecture2}. Each internal service relies on an authentication token that encodes the client’s scopes and roles, ensuring that actions are authorized correctly.  Tokens are issued by an OAuth 2.0 service that is bound to the internal Kerberos authentication and LDAP directory, which stores user roles and group memberships.  The architecture also supports external OAuth 2.0/OIDC providers for obtaining user attributes.

\subsection {FOXDEN Services}
\label{sec:services}
All services share a common Go‑based library (see Section~\ref{sec:implementation}) that provides routing, middleware, authentication, and data‑access primitives, guaranteeing a uniform footprint and behaviour across the platform.

\begin{itemize}
    \item The \textbf{Frontend Service} provides convenient web and command-line interfaces to forward requests and search queries to the other services and organizes the returned results.
    \item The \textbf{Authn/Authz Service} issues OAuth‑compatible tokens and enforces their scopes. Users authenticate via Kerberos tickets, which are exchanged for FOXDEN access tokens. Each token carries a set of scopes that restrict the holder to specific operations (\emph{e.g.}, read, write, admin), thereby providing fine‑grained, auditable access control across all FOXDEN components.
    \item The \textbf{Metadata Service}, \textbf{Provenance Service}, and \textbf{SpecScan Service} record and organize the metadata records described in Section~\ref{sec:record-types}, making them available for fast retrieval by downstream data analysis pipelines that consume the corresponding datasets and scans.
    \item The \textbf{Data Management Service} guarantees the availability of raw and derived datasets throughout their entire lifecycles. It handles ingestion, persistent storage, archival, and retrieval, enforcing access policies defined by the Authn/Authz Service.
    \item The \textbf{Data Discovery Service} traverses the other services above to aggregate records associated with a given DID. It presents users with a unified query language, and it can be extended to additional services and metadata sources in the future.
    \item The \textbf{DOI Publication Service} is a proxy server that connects users to DOI providers such as DataCite \cite{DataCite}, Materials Commons \cite{MaterialsCommons}, and Zenodo \cite{Zenodo}. It allows users to mint DOIs for chains of metadata records, as shown in Figure~\ref{fig:info-layout}.
    \item The \textbf{Sync Service} synchronizes instances of FOXDEN installed at different locations. It ensures that metadata records and access control information remain consistent across geographically distributed deployments.
\end{itemize}

All data service endpoints are REST‑compliant and implement the conventional HTTP verbs: POST for creating resources, PUT for updating them, DELETE for removal, and GET for retrieval.  Each service can be horizontally scaled using Kubernetes deployments to meet the desired throughput.

Because the authentication layer can be bypassed with a simple configuration switch, FOXDEN can easily be installed in small research labs, as well as large facilities, and it can run either on local or central resources, such as a user's laptop or a dedicated Kubernetes cluster.

In the future, we plan to augment the above services with a set of collaboration services, as shown in Figure~\ref{fig:foxden_architecture2}. The \textbf{AI/ML Service} will be a dedicated repository for AI/ML models with an inference engine that will act as a proxy to common platforms such as TensorFlow, PyTorch, Keras, and Scikit-learn. This service, modeled on an existing design for high-energy physics~\cite{mlaas}, will offer versioned storage, model metadata, and access control, enabling reproducible AI/ML workflows.
The \textbf{Notebook Service} will simplify data analysis by giving users a Jupyter-like interface for writing code modules that are inserted into predefined workflows and also deposited in a code repository for future dissemination.
The \textbf{Visualization Service} will serve as an interface for easily accessing and managing data visualization dashboards developed in collaboration with the National Science Data Fabric~\cite{nsdf-dashboards}. 
The \textbf{Analysis Service} will provide an interface to the X-ray Imaging of Microstructures Gateway (XIMG)~\cite{ximg-website, ximg-poster, ximg-pasc}, which is an instance of the Galaxy science gateway~\cite{Galaxy}, developed to preserve and enable reuse of data analysis workflows for materials science.

\subsection{Service Repositories}
The FOXDEN services described above are backed by repositories consisting of relational and document databases. To keep the codebase flexible, we introduced generic Go interfaces \cite{edwards_letsgo} that abstract the underlying database APIs, allowing backends to be swapped without modifying the service implementation. For the relational databases, FOXDEN supports any choice, from embedded SQLite~\cite{sqlite,go-sqlite3} engines to enterprise‑grade systems such as Oracle. In our deployments, we have used MySQL for the production environment at CHESS and SQLite for off‑site installations. For document databases, we use MongoDB.

\subsection{Storage Abstraction Layer}
FOXDEN's storage abstraction layer provides a uniform interface to diverse resources, such as institutional filesystems and cloud object storage.
Currently, we support on‑site access at CHESS through Globus~\cite{globus-foster,globus-acm} APIs as well as an S3‑compatible interface. The architecture is extensible to future federated storage solutions (\emph{e.g.}, iRODS~\cite{iRODs}, Ceph~\cite{ceph}, or custom object stores).

\section{Sustainable Implementation}
\label{sec:implementation}

With FOXDEN, sustainability is not an afterthought but a cornerstone of the design. Each architectural feature has been carefully considered within a long-term strategy of prioritizing adaptability and growth. This focus on sustainability extends to FOXDEN's implementation, from the choice of programming language to operational considerations.

\subsection{Go Programming Language}
We have chosen Go \cite{GoLang} as the most suitable programming language for FOXDEN implementation. Unlike other high-level languages like Python and C/C++, Go combines a performant compiled runtime with a comprehensive standard library that covers HTTP handling, templating, and JSON processing out of the box, eliminating the need for external dependencies. Go's syntax is deliberately minimal and approachable—comparable to Python—while built‑in concurrency primitives (goroutines and channels) simplify parallel workloads. Moreover, Go produces a single static executable, which streamlines deployment, reduces dependency‑related failures, and lowers operational costs across our heterogeneous cluster. These attributes collectively make Go the optimal choice \cite{GoLang4AI} for delivering a robust, maintainable, and efficient web service architecture for FOXDEN.

Although the FOXDEN services\footnote{We consolidated the core functionality into a dedicated golib library \cite{foxden_golib} that is reused by all FOXDEN services. For example, server‑side components such as routing, middleware, rate‑limiting, database APIs, lexicon rules, and common utilities are shared across every service.} themselves are developed in Go, their implementation of the HTTP protocol allows them to be invoked from any programming language. For instance, Python clients can send requests to FOXDEN HTTP endpoints to inject, retrieve, and manage metadata.

\subsection{Database Scalability}
\label{sec:databases}
To ensure the scalability of FOXDEN's infrastructure, we organize its metadata records into separate databases to account for differences between structured and unstructured metadata, as well as between high and low volumes of records. Generic Metadata records are unstructured, so they are stored in document databases that provide the flexibility to handle multiple schemas at the expense of performance. However, Provenance and SpecScan records are highly structured with uniform schema across all datasets and beamlines. With these records, the flexibility of document databases is not required, so they are stored in more performant relational databases instead.

Furthermore, SpecScan records outnumber Provenance records by at least an order of magnitude. To optimize search and retrieval for the low-volume metadata, we store SpecScan and Provenance records in separate relational databases. Attachments and \emph{ad hoc} records without defined schemas are likewise stored separately from the generic Metadata. Table~\ref{tab:databases} summarizes the characteristics of Metadata, Provenance, and SpecScan records.

\begin{table}[ht]
\centering
\caption{FOXDEN database organization. We store coarse-grained Metadata records in a document database and fine-grained Provenance and SpecScan records with standardized schemas in relational databases.}
\begin{tabular}{cccc}
\hline\hline
Type & Volume & Schema & Database \\
\hline
Generic Metadata & Low & Variable & Document \\
Provenance & Medium & Fixed & Relational \\
SpecScan & High & Fixed & Relational \\
\hline\hline
\end{tabular}
\label{tab:databases}
\end{table}

\subsection{Operational Resources}
By developing services in a uniform manner~\cite{foxden_repos}, we conserve effort and minimize the final binary footprint.
Each service compiles to an executable of roughly 30 MiB. The current production deployment of a dozen services comfortably fits on a single node with 2 CPU cores and 8 GiB RAM\footnote{In contrast, a comparable Python web application usually requires a full interpreter, virtual‑environment dependencies, and often additional runtime libraries. In practice, this means the hardware footprint described above is the bare minimum needed to run even a single Python service.}.
Identical nodes are used to host development and demonstration instances which run 24 individual services combined.

By leveraging modular design from the outset, FOXDEN resources can be dynamically scaled in response to growth in data volumes and user demand, thus avoiding costly retrofits in the future. This scaling can occur both vertically—handling higher throughput via concurrency on a single node—and horizontally—adding nodes using orchestration frameworks like Kubernetes to accommodate more users or larger datasets.
We provide Kubernetes manifest files for FOXDEN services, although the current deployments do not use them.

\section{Usage}
FOXDEN is designed for both interactive and scripted workflows.
For interactive use, FOXDEN's web interface offers an intuitive, graphical means of manually performing basic operations, such as injecting, retrieving, updating, and publishing Metadata and Provenance records. SpecScan records can only be retrieved and published through the web interface. Search queries are formulated using the MongoDB query language~\cite{mongodb-query} and are composed in JSON format. A query builder with dropdown menus for selecting schema attributes assists users with the JSON syntax. 
To facilitate navigation through retrieved records, linked records are visualized as tree structures (similar to that shown in Figure~\ref{fig:info-layout}) with clickable nodes.

Users are authorized to view and modify only those records that they or their research group have created. Research group membership is defined by Unix groups and is queried through the Lightweight Directory Access Protocol (LDAP).

For expert users and scripted workflows, we provide an HTTP API as well as a powerful command-line interface (CLI) that can be used for bulk record creation and automation. The API and CLI both allow for granular control over basic operations as well as advanced functionality, such as managing authentication tokens and deleting records. In particular, the only way to create and modify SpecScan records is through the API and CLI, not the web interface.

The FOXDEN HTTP API may be invoked from the command line, as in the following example:
\begin{verbatim}
    # Add a new Metadata record
    curl -v -X POST -H "Content-type: application/json" \
    -H "Authorization: Bearer $token" \
    -d@./record.json http://foxden.example.url:8300
\end{verbatim}

The FOXDEN CLI performs identical actions but with a standard, intuitive syntax for quick adoption:
\begin{verbatim}
    # Generic command pattern
    foxden <command> <operation>

    # Example commands
    foxden meta add file.json           # Add a new Metadata record
    foxden prov ls parents --did=$DID   # Find a dataset's parents through its Provenance
\end{verbatim}
A full list of commands is available in the FOXDEN documentation~\cite{foxden_docs}.

When a record is injected into FOXDEN, its contents are validated against the specified schema, and an error is thrown when mismatches are detected.
After injection, a record can be revised without rewriting the entire object; FOXDEN tracks each revision as a new Provenance node, thus preserving the entire history of the record.

In addition, partially completed Metadata records are treated on the same footing as completed records. Before the start of an experiment, users may submit template records with placeholder fields that are later filled in as the experiment progresses. This staged approach fosters consistent metadata practices by reducing the bottlenecks that might discourage real-time recording of metadata.

\section{Research Integration}

In this Section, we describe the deployment of FOXDEN at two large research facilities that each serve large user pools spanning multiple disciplines: CHESS and the National High Magnetic Field Laboratory (MagLab). In the past two decades, both facilities have seen dramatic changes in the way data are acquired and analyzed. Due to improvements in instrumentation and data acquisition systems, datasets have grown by orders of magnitude in terms of number, volume, and complexity. Concurrently, requirements and responsibilities around data sharing have evolved. It is now incumbent upon researchers to ensure that the provenance, integrity, and security of datasets are maintained throughout the data lifecycle, from generation to publication. It is also essential that datasets are FAIR, with a particular emphasis on ensuring the reproducibility of experimental results. The nature of CHESS and the MagLab as publicly-funded research institutions requires careful stewardship of the large volume of high quality research data they generate, to ensure the greatest possible return on investment for taxpayers. In the era of rapid change brought on by the adoption of AI in scientific research, this means ensuring that datasets are AI-ready (which is enabled by making them FAIR) so that new methods can be applied and/or developed, yielding increased scientific value and benefit to society.

At CHESS and other synchrotron facilities, traditional methods of data management are increasingly insufficient to keep pace with the vast volumes of data generated. Historically, facility users took raw data collected at the beamline back to their home institutions for processing and analysis on local compute resources, often transporting raw data on physical hard drives. As data volumes and analysis workflows have increased in size and complexity, users are increasingly relying on facility resources for these tasks. Simultaneously, the way researchers create and maintain records of their experimental work has evolved. Although the increased use of electronic record-keeping has the potential benefit of facilitating information sharing compared to physical lab notebooks, in practice, the variety of platforms (\emph{e.g.}, Google docs, locally saved text files and spreadsheets, individual email accounts, etc.) has resulted in a fragmented metadata landscape. 

Similarly, at the MagLab, new data and metadata management challenges confront MagLab users and support staff. Historically, MagLab facilities have taken a \textit{laissez faire} approach to data management that focuses on minimizing barriers to entry and maximizing convenience and utility for users. Accordingly, data are often collected by (or handed off immediately to) users who apply their preferred data analysis methods with little or no involvement from MagLab support staff. Metadata acquired along with the data are often limited, tied up in proprietary file formats, or nonexistent. While the MagLab has achieved multiple decades of extraordinary productivity, this approach directly conflicts with the goals of ensuring that provenance is tracked, integrity is maintained, and datasets are FAIR, all of which benefit greatly from intervention at the time of acquisition. In addition, existing data acquisition and analysis pipelines vary across the MagLab's facilities and rely on \emph{ad hoc} combinations of vendor-provided and/or custom-made software and hardware. Many do not automatically acquire detailed metadata with sufficient attributes to make datasets FAIR, stymieing efforts towards application of AI. There is also no method in place to ensure that datasets and any associated metadata are catalogued to enable discoverability and reuse.

The suite of services offered by FOXDEN provides the ideal foundation to build the cyberinfrastructure needed to tackle these challenges.
FOXDEN imposes minimal requirements on the operating systems and hardware architecture of its host computers, which makes it suitable for environments with limited computing resources. The Go language's cross‑compilation ability allows FOXDEN to be deployed as a single static executable on any operating system. Instead of managed database services, the backend repositories may be implemented as embedded databases such as SQLite~\cite{sqlite, go-sqlite3} and Badger~\cite{badger}. 
Finally, the inherent flexibility of FOXDEN schemas makes the platform adaptable to a wide variety of research settings, serving different domain-science communities with distinct conventions and needs. 

\subsection{CHESS}

CHESS is one of only six high-energy (\emph{i.e.}, particle beam energy $\geq$ 6 GeV) storage ring light sources in the world, providing high-flux, high-energy x-rays for experiments probing the structure and dynamics of physical systems down to the atomic scale. Currently, CHESS operates seven beamlines, with two additional beamlines under construction. CHESS serves an exceptionally broad community of users from fields including biology, chemistry, materials science, engineering, and cultural heritage. 

\begin{figure}[ht]
\begin{center}
\includegraphics[width=0.7\linewidth]{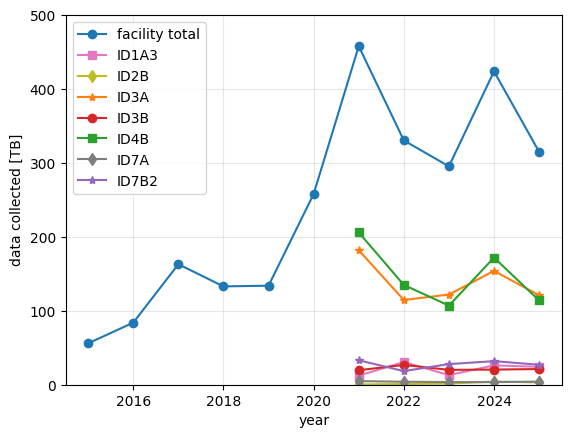}
\caption{\label{fig:chess-data-vol}Raw data collected per year at CHESS. Per-beamline data volumes are shown starting in 2021, the first year for which this information is available.}
\end{center}
\end{figure}

As mentioned in Section \ref{sec:intro}, the volume of raw and processed data collected at CHESS varies widely by beamline. As shown in Figure \ref{fig:chess-data-vol}, the ID3A~\cite{chess-id3a}
and ID4B~\cite{chess-id4b}
beamlines are responsible for a significant fraction of the total facility data volume, each collecting more than 100 TB of raw data per year. Prompted by the challenges inherent in implementing FAIR data practices on these large data volumes, both beamlines have served as early-stage testing sites for FOXDEN services.

Both ID3A and ID4B specialize in diffraction-based techniques involving the collection of many thousands of images for each sample on large-area ($\geq$ 6 megapixels with $\geq$ 14 bits/pixel) detectors. At ID3A, also known as the Forming and Shaping Technology beamline (FAST), the most data-intensive technique is far-field 3D x-ray diffraction (3DXRD, also referred to as high energy diffraction microscopy or HEDM), which uses diffraction data to reconstruct the centroid, elastic strain tensor, and crystallographic orientation of individual grains within millimeter-scale volumes of polycrystalline samples \cite{bernier2020}. These measurements, when conducted \emph{in situ} during thermomechanical loading, help materials researchers uncover the relationships between a material's microstructure and its mechanical properties and performance. On the other hand, the Q-Mapping for Quantum Materials beamline (QM2) at ID4B is designed for high-throughput characterization of quantum materials in reciprocal space, for both single crystals and grazing-incidence thin films ($\geq$ 3 nm) \cite{Sarker2026QM2}. The beamline collects comprehensive data on thousands of Brillouin zones (BZs), enabling users to perform 120 million distinct Q-point measurements each second (6 million pixels at 20 Hz frame rates) to capture weak signals, uncovering intertwined quantum correlations of spins, charges, and orbitals from high to low temperatures and under a variety of perturbations, spanning multi-dimensional phase diagrams.

FOXDEN implementation at ID3A and ID4B offers a clear illustration of the need to accommodate a variety of user requirements, even within a single facility. The Metadata schemas for ID3A and ID4B contain some common key/value pairs (\emph{e.g.}, beam energy, list of experimenters), but many other key/value pairs
are dictated by the needs of the individual beamline's user community and not shared in common (\emph{e.g.}, mechanical test and load frame type at ID3A \emph{vs.} energies of sample absorption edges at ID4B). 

At both beamlines, Metadata record creation is orchestrated by SPEC, typically when a user begins data collection on a new sample. Customized scripts combine ``beamtime-level" information, populated by facility staff at the beginning of a user beamtime, with ``sample-level" information provided by the user. Currently, users may supply this information either through SPEC command-line prompts or through a separate JSON file. Once the complete Metadata record is assembled, SPEC triggers the execution of a bash script that submits the record to the FOXDEN Metadata service. SPEC displays a message for the user indicating whether the record submission was successful, and providing the DID if so. Alternatively, the SPEC processes that generate and submit Metadata records can be disabled by beamline staff, for cases where it is preferable to use the FOXDEN web UI or CLI to create and submit records directly.

Early in the beamline testing process, we identified a need for and implemented two critical usability features: 1) the ability for data collection to proceed without interruption even if errors arise during the generation or submission of a Metadata record, and 2) the ability for users to amend records after submission to correct errors or oversights. Both have been essential to enabling us to pilot FOXDEN services during experiments run by external CHESS users, whose first priority is to successfully execute their x-ray measurements in the limited amount of beamtime allocated to them. In turn, feedback from these users has been essential to improving the FOXDEN UI and setting priorities for new feature development.

After data collection, CHESS users process their raw data using a variety of software packages, including CHESS-maintained code, commercially available software, open-source community-maintained software, and bespoke user-developed scripts. To accommodate this exceptionally broad and constantly evolving software landscape, FOXDEN provides the ability to upload a ``user" Metadata record including the application used to process the parent dataset, a list of input files, a list of output files, and an arbitrary number of user-defined key/value pairs intended to hold whatever information the user deems critical for documenting their data processing workflow. As described in Section \ref{sec:research-workflow} and illustrated in Figure \ref{fig:foxden-workflow}, the relationship between this record and its parent is preserved in the form of a Provenance record. Additional records can be created as processed data is analyzed using additional tools, or as raw data is re-processed using different tools or parameters. Workflows that are constructed with a data analysis framework known as the CHESS Analysis Pipeline (see Appendix~\ref{sec:chap}) can leverage built-in modules that read and write FOXDEN records directly.

Finally, these curated metadata records can be published through external DOI providers, enabling long-term preservation, discovery, and citation of experimental datasets. In recent years, research conducted at ID4B has led to multiple peer-reviewed publications using metadata records generated through FOXDEN, either by exporting metadata records to external repositories such as Zenodo  \cite{Turkiewicz2025, Salinas2025, GomezAlvarado2025}  or initiating DOI registration in Materials Commons  \cite{Yang2026} directly through FOXDEN's
DOI Publication Service.

To prepare curated, publication-ready ID4B metadata records, beamline scientists collaborate closely with the user group's Principal Investigator (PI) to obtain approval for data sharing and to ensure that comprehensive Metadata and Provenance information are validated throughout the experimental lifecycle. FOXDEN streamlines the metadata publication workflow by providing an integrated interface for selecting a preferred repository, including Materials Commons or other DataCite-supported providers. Once the publication-ready metadata record is created, FOXDEN generates and displays key information, including the DOI, provider name, provider-specific record identifier, data access link, and publication date. Additionally, beamline scientists or users establish and manage the necessary repository accounts and permissions required by the chosen provider, such as Materials Commons or Zenodo. Following DOI registration, the persistent dataset link is shared with collaborators and cited in subsequent journal articles \cite{GomezAlvarado2026, CapaSalinas2026, Turkiewicz2026}.
By integrating metadata management, provenance tracking, and DOI registration, the FOXDEN framework facilitates reproducible and data-driven scientific research while improving the discoverability and long-term accessibility of experimental data.

\subsection{National High Magnetic Field Laboratory}

The National High Magnetic Field Laboratory (MagLab) is an NSF major facility that focuses on the use of high magnetic fields for scientific research across a broad variety of disciplines. Its seven user facilities focus on different instrumental techniques and applications and serve diverse user communities in chemistry, biochemistry, condensed matter physics and materials science, among others. Two additional research facilities focus on the development of novel superconducting materials and high field magnet systems, making the MagLab the world leader in both development and application of these technologies. Many of the instruments at the MagLab are world-unique and record-breaking, offering capabilities to users that cannot be replicated elsewhere.

\subsubsection{ICR Facility Implementation}

The prototype implementation of FOXDEN-based cyberinfrastructure is in the MagLab's Ion Cyclotron Resonance (ICR) user facility, which focuses on the application of Fourier transform-ion cyclotron resonance mass spectrometry (FT-ICR MS) to characterization of complex mixtures of analytes extracted from biological, petroleum, and environmental samples. FT-ICR MS is the highest performing mass spectrometry (MS) technique but requires significant investments to enable, \emph{i.e.}, high field magnets that require extensive infrastructure and expertise to design, build, and operate. This is a primary motivation behind making the ICR facility's instruments, such as its flagship 14.5 tesla (T)~\cite{schaub_high-performance_2008} and 21 T FT-ICR mass spectrometers~\cite{hendrickson_21_2015}, available as a national resource for qualified user projects. The facility's instruments have been continuously utilized and regularly updated throughout the years and have generated numerous high quality datasets which remain locally archived.

The ICR facility faces some of the greatest data and metadata management challenges of all MagLab user facilities. The rate at which high resolution MS data can be generated has increased significantly in recent years due to a rise in the number of time-resolved liquid chromatography-mass spectrometry (LC-MS) and spatially-resolved mass spectrometry imaging (MSI) experiments, both of which can generate data at a rate of hundreds of terabytes per year, similar to beamline facilities such as CHESS. ICR instruments and experiments are complex, and a great deal of experimental metadata must be captured to effectively enable reproducibility, integrity, and AI-based workflows.

The facility's flagships instruments, 14.5 T and 21 T FT-ICR mass spectrometers, are hybrid systems built on front end hardware (Orbitrap Eclipse mass spectrometers, Thermo Fisher Scientific, Waltham, MA) and software systems (Xcalibur, id.) with extensive custom modifications. The primary data format for these systems is a proprietary .Raw file format containing mass spectrometry data and metadata pertaining to instrumental and data acquisition parameters. To apply to FOXDEN to ICR facility data, staff at the MagLab's Center for FAIR and Open Science developed a prototype software interface called FOXDelegate. This lightweight GUI-based program utilizes FOXDEN services to generate Metadata records from data collected by ICR facility instrumentation. FOXDelegate uses software provided by the manufacturer to programmatically extract metadata, then formats it according to custom schemas and uploads Metadata records to FOXDEN either at the time of file generation or afterwards to allow historical dataset metadata to be catalogued.

FOXDEN's composable schemas were easily adapted to address the needs of the ICR facility, and \textit{ad hoc} schema extensions provided the capability to store the extracted metadata in its original form or using metadata standards such as DataCite, Schema.org, and Croissant that facilitate discoverability and machine-actionability. During testing, LLMs were able to easily read, compare, and describe data acquisition and analysis methods used to acquire FT-ICR datasets from the detailed metadata available in FOXDEN Metadata records generated from .Raw files. These capabilities will be extended to ICR facility instrumentation and software provided by other vendors (\emph{e.g.}, Bruker Corporation, Billerica, MA) and using custom data acquisition software (\emph{e.g.}, Predator)~\cite{blakney_predator_2011}, so that the same unified metadata formats can be broadly applied to facilitate integration of datasets collected by different methods. Also, additional schemas will be generated to enable improved metadata collection for particular application areas within the ICR facility (\emph{e.g.}, petroleum analysis \emph{vs.} proteomics).

New cyberinfrastructure built on FOXDEN services will allow ICR facility users and support staff to quickly and seamlessly automate the generation and cataloguing of detailed metadata for every file and dataset collected using FT-ICR MS instruments. This will ensure the chain of provenance of every research output is established at the moment of generation, build the facility's internal database of detailed experimental metadata from both new and historical datasets, and allow users to immediately apply AI-based workflows to the data they acquire thanks to metadata formatting standards that provide facile machine actionability with FOXDEN services as the orchestrator. Access to datasets and metadata records can be tightly controlled through the use of appropriately scoped authentication tokens, ensuring user intellectual property is protected. The ICR facility also has the opportunity to implement a data storage strategy similar to that of CHESS wherein large datasets and lightweight metadata records utilize separate storage infrastructures for maximum efficiency. In time, innovations in the ICR facility will be rolled out to other MagLab facilities, bringing the benefits of FAIR and AI-ready datasets to the broader user community and impacting additional fields of research such as condensed matter physics and materials science. 

In the latter field, efforts such as the Materials Genome Initiative, a federal multi-agency initiative to accelerate the pace of discovery and design of new materials for a broad variety of applications using advanced computational techniques \cite{de_pablo_materials_2014}, have led to new materials informatics approaches \cite{shahzad_accelerating_2024} to address critical societal challenges \cite{stier_materials_2024}. The flexibility of FOXDEN services and templates to adapt to metadata standards that are relevant for a specific research discipline will allow it to be leveraged for these new materials informatics approaches by integrating into existing FAIR materials data architectures \cite{aggour_semantics-enabled_2024}. FOXDEN already features an integration with the Materials Commons repository \cite{puchala_materials_2016}, providing immediate access to a database relevant to many MagLab users in the materials science field. Additional integrations could extend functionality to additional discipline-specific databases.

\section{Summary and Future Directions}
\label{sec:future}

FOXDEN is a suite of lightweight, modular data services that encapsulates a domain-agnostic approach to recording and disseminating scientific metadata. Because FOXDEN services can be invoked via a generic HTTP API, metadata can be injected by any data acquisition program, then retrieved and augmented in downstream data analysis workflows. Therefore, FOXDEN functions as a structured electronic lab notebook, enabling the seamless flow of information across disparate systems and matching the shape of any type of research workflow. In addition, FOXDEN promotes FAIR data principles by allowing DOIs to be minted for chains of metadata records, thus making them accessible through public repositories. 

We have deployed a basic set of FOXDEN services at CHESS and the MagLab, with over 10000 records collected to date, and we are refining the user interface in response to researcher feedback. In addition to the improvements described above in Sections~\ref{sec:generic-metadata} and~\ref{sec:services}, upcoming efforts will focus on growing the services layer and underlying infrastructure, including:
\begin{itemize}
    \item {\bf Federated data management}: Extending FOXDEN into a truly distributed ecosystem will enable collection of research artifacts across multiple sites. This work includes implementing fine‑grained access controls (\emph{e.g.}, per‑file or per‑slice permissions) and support for streamed data delivery, which is essential for continuous AI training pipelines.
    \item {\bf AI integrations}: Augmenting FOXDEN with a local vector database that stores document embeddings from experimental data allows for an LLM chatbot interface that retrieves data and metadata in response to natural language prompts. Such an LLM can also be woven into the knowledge graph described in Section~\ref{sec:info-layout}, enabling cross-facility searches and multimodal queries that combine text with example dataset content. With this interface, AI agents may autonomously discover datasets and deliver them to other services for modeling, inference, or analysis. 
    \item {\bf Data fabric integration}: Effective research data management requires the implementation of a data fabric, a comprehensive data management architecture made up of components that integrate, connect, and govern access to multidisciplinary scientific data and metadata from disparate sources while providing the necessary semantic context for machine actionability \cite{blohm_data_2024}. Constructing and managing such an architecture requires the development and continual revision of cyberinfrastructure components to carry out various tasks, \emph{e.g.}, metadata extraction, formatting, and cataloguing. The lightweight and flexible nature of FOXDEN services makes them ideal for incorporation as key components of new CI for this purpose.
\end{itemize}
These future directions are built on the foundation of FOXDEN's FAIR infrastructure, which enhances existing scientific data pipelines with scalable metadata capabilities. Thus, FOXDEN bridges the gap between exploratory and automated workflows. It captures the critical metadata that unlocks bespoke datasets, allowing them to be reused and leveraged manyfold in comprehensive, autonomous, AI-driven research.

\section*{Acknowledgements} We thank the following individuals for invaluable assistance and helpful discussions: Khaled Alharbi, John Allison, Devin Bougie, Joel Brock, Amlan Das, Jun Young Peter Ko, Wendy Kozlowski, Matthew Miller, Valerio Pascucci, Giorgio Scorzelli, Glenn Tarcea, Estella Yee.

This work is based on research conducted at the Center for High-Energy X-ray Sciences (CHEXS), which is supported by the National Science Foundation (BIO, ENG and MPS Directorates) under award DMR-2342336, and the Macromolecular Diffraction at CHESS (MacCHESS) facility, which is supported by award P30GM124166 from the National Institute of General Medical Sciences and the National Institutes of Health. This material is based on research sponsored by AFRL under agreement number FA8650-22-2-5200. The U.S. Government is authorized to reproduce and distribute reprints for Governmental purposes notwithstanding any copyright notation thereon.

The National High Magnetic Field Laboratory is supported by National Science Foundation Division of Materials Research and Division of Chemistry through DMR-2128556 and the State of Florida.

\bibliography{references}
% I prefer to use the IEEE bibliography style. 
% That's  NOT required by the NSF guidelines. 
% Feel Free to use whatever style you prefer
%\bibliographystyle{IEEEtran}
%\bibliographystyle{unsrtnat}
\bibliographystyle{unsrturl}
%\printbibliography

\newpage
\appendix
\section{CHESS Analysis Pipeline}
\label{sec:chap}
\begin{figure}[ht]
\begin{center}
\includegraphics[width=0.3\linewidth]{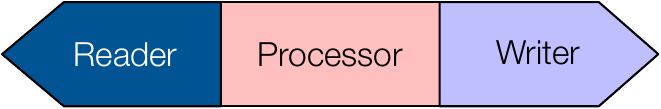}
\caption{\label{fig:chap-base}A CHAP \texttt{Pipeline} component composed of \texttt{Reader}, \texttt{Processor}, and \texttt{Writer} base classes.}
\end{center}
\end{figure}

The CHESS Analysis Pipeline (CHAP)~\cite{CHAP} is an object-oriented Python framework for refactoring monolithic data analysis programs into modular pipelines composed of interchangeable, reusable code components. The fundamental components of CHAP are the base classes \texttt{Reader}, \texttt{Processor}, and \texttt{Writer}, which can be assembled into \texttt{Pipeline} components, as shown in Figure~\ref{fig:chap-base}. 

These base classes isolate a \texttt{Pipeline}'s I/O functions from its data analysis algorithms, allowing a single algorithm to accept multiple data formats.
\begin{itemize}
\item \texttt{Reader} and \texttt{Writer} classes handle data input and output, respectively. They encapsulate CHESS-specific logistics, such as file operations and data format conversions, and they validate their data with the Pydantic~\cite{pydantic} library. Inherited subclasses are defined for specific file types, \emph{e.g.} \texttt{H5Reader} and \texttt{NexusWriter}. A single \texttt{Pipeline} component may contain multiple \texttt{Reader}s or \texttt{Writer}s.
\item The \texttt{Processor} class encapsulates data analysis algorithms. Because \texttt{Processor}s are independent of CHESS infrastructure, researchers can easily contribute \texttt{Processor} subclasses with bespoke data analysis code. \texttt{Reader}s and \texttt{Writer}s pass data into and out of a \texttt{Processor} via \texttt{PipelineData} containers. Multiple \texttt{Processor}s can be chained together within a single \texttt{Pipeline} component.
\end{itemize}

\begin{figure}[ht]
\begin{center}
\includegraphics[width=0.99\linewidth]{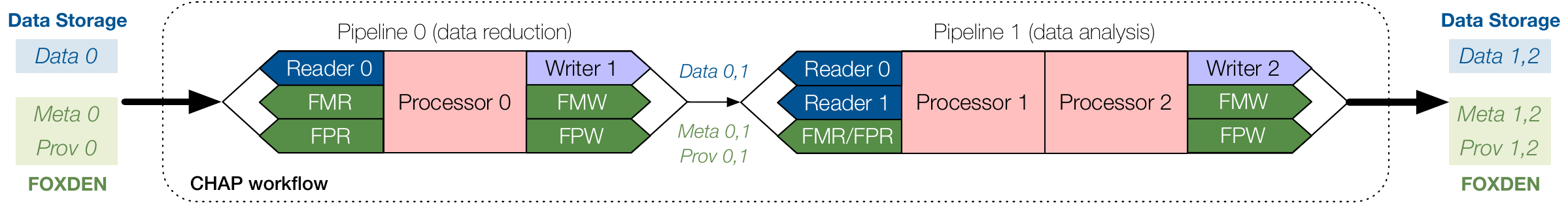}
\caption{\label{fig:chap-workflow} A schematic CHAP workflow composed of two \texttt{Pipeline} components that perform successive stages of data reduction and analysis. FMR and FPR denote the \texttt{FoxdenMetadataReader} and \texttt{FoxdenProvenanceReader} classes, and FMW and FPW denote \texttt{FoxdenMetadataWriter} and \texttt{FoxdenProvenanceWriter}, respectively.}
\end{center}
\end{figure}

Workflows are defined by CHAP configuration files written in YAML. Each configuration file may contain one or more \texttt{Pipeline} components that can be executed individually or all at once (sequentially or in parallel). Figure~\ref{fig:chap-workflow} shows a schematic workflow with a series of two \texttt{Pipeline} components linked together in a single configuration file.

CHAP \texttt{Processor}s typically need metadata, which can be provided by user-generated input files and read by suitable CHAP \texttt{Reader}s. Alternatively, CHAP can retrieve metadata directly from FOXDEN (using the Metadata and Provenance Services), relieving the user from having to manually provide metadata with the associated risk of errors or incomplete information. To streamline this process, we provide a \texttt{FoxdenMetadataReader} and \texttt{FoxdenProvenanceReader} (derived classes of \texttt{Reader}) that call FOXDEN's HTTP APIs to retrieve the relevant records from within a \texttt{Pipeline}, as shown in Figure~\ref{fig:chap-workflow}. Similarly, \texttt{FoxdenMetadataWriter} and \texttt{FoxdenProvenanceWriter} (derived classes of \texttt{Writer}) inject FOXDEN records for the new datasets produced by CHAP \texttt{Processor}s.

Using FOXDEN in a CHAP workflow, as illustrated in Figure~\ref{fig:chap-workflow} for example, is as easy as adding a \texttt{FoxdenMetadataReader} and \texttt{FoxdenProvenanceReader} to the CHAP YAML configuration file as inputs to the first CHAP \texttt{Processor}. Each of these \texttt{Reader}s is supplied with a configuration specifying the appropriate FOXDEN service URL as well as the input dataset DID (or, alternatively, a FOXDEN Data Discovery Service query for the input dataset). CHAP \texttt{Processor}s can then access any attributes available in the Metadata and Provenance records via a call such as \texttt{record.get(`sample\_name')}, where \texttt{sample\_name} is an example attribute defined in the corresponding schema.

CHAP \texttt{Processor}-specific Metadata is automatically appended to the \texttt{PipelineData} list and can either be passed on to the next \texttt{Pipeline} component or written with the \texttt{FoxdenMetadataWriter} to the FOXDEN Metadata Service. The appropriate Provenance records and the correct parent-child relationships are created automatically and can be written with a \texttt{FoxdenProvenanceWriter} to the FOXDEN Provenance Service.

The CHAP API comes with a \texttt{run()} method for each \texttt{Reader}, \texttt{Writer} and \texttt{Processor}, which allows users to reuse any CHAP component straightforwardly in standalone python (Jupyter) notebooks or scripts.

The integration of CHAP with FOXDEN demonstrates the benefits of using machine-readable metadata to streamline the configuration of data pipelines: human error is reduced, and as a result, the pipelines can be easily scaled up and automated.

\end{document}